# Possible cool prominence materials detected within interplanetary small magnetic flux ropes

J. M. Wang, H. Q. Feng (fenghq9921@163.com), H. B. Li, A. K. Zhao, Z. J. Tian, G. Q. Zhao, Y. Zhao, Q. Liu

Institute of space physics, Luoyang Normal University, Luoyang, China

**Abstract**

Previous studies indicate that interplanetary small magnetic flux ropes (SMFRs) are manifestations of microflare-associated small coronal mass ejections (CMEs), and the hot material with high charge states heated by related microflares are found in SMFRs. Ordinary CMEs are frequently associated with prominence eruptions, and cool prominence materials are found within some magnetic clouds (MCs). Therefore, the predicted small CMEs may also be frequently associated with small prominence eruptions. In this work, we aim to search for cool prominence materials within SMFRs. We examined all the $O^{5+}$ and $Fe^{6+}$ fraction data obtained by the *Advanced Composition Explorer* spacecraft during 1998 to 2008 and found that 13 SMFRs might exhibit low-charge-state signatures of unusual $O^{5+}$ and/or $Fe^{6+}$ abundances. One of the 13 SMFRs also exhibited signatures of high ionic charge states. We also reported a SMFR with high $Fe^{6+}$ fraction, but the values of $Fe^{6+}$ is a little lower than the threshold defining unusual $Fe^{6+}$. However, the SDO/AIA observations confirmed that the progenitor CME of this SMFR is associated with a small eruptive prominence, and the observations also supported the prominence materials were embedded in the CME. These observations are at the edge of the capabilities of ACE/SWICS and it cannot be ruled out that they are solely caused by instrumental effects. If these observations are real, they provide new evidence for the conjecture that SMFRs are small-scale MCs but also imply that the connected small CMEs could be associated with flares and prominence eruptions.

## 1. Introduction

In interplanetary space, two kinds of magnetic flux ropes (MFRs) are often observed by spacecraft near the Earth's orbit (Moldwin et al. 2000; Cartwright & Moldwin 2008; Feng et al. 2007, 2008, 2015; Feng & Wang 2015). One kind is magnetic clouds (MCs), which are a subset of interplanetary coronal mass ejections (ICMEs); their diameters are approximately from 0.20 AU to 0.40 AU, and their durations are usually between 12 and tens of hours (Burlaga et al. 1981; Lepping et al. 1990). The other kind is small MFRs (SMFRs); their diameters are usually less than 0.20 AU, and their durations are usually between tens of minutes and several hours (Feng et al. 2008; 2015). MCs and SMFRs have two essential characters: enhanced magnetic field strength and smooth rotation of magnetic field vector; they can be fitted with cylindrically symmetric flux ropes (Moldwin et al. 2000; Feng et al. 2007, 2008, 2010b). Coronal mass ejections (CMEs) were frequently correlated with other solar activities such as prominence eruptions and solar flares (Webb et al. 1976; Munro et al. 1979). Munro et al. (1979) reported that nearly 70% of CMEs are correlated with prominence eruptions, and 40% are correlated with flares. Correspondingly, many observations of prominence plasma materials within MCs (e.g. Burlaga et al. 1998; Skoug et al. 1999; Gloeckler et al. 1998; Lepri & Zurbuchen, 2010; Yao et al. 2010; Sharma et al. 2013; Wang et al. 2018), which are the interplanetary manifestations of CMEs (Démoulin 2008), were reported. High mean Fe charge states ($<Q_{Fe}> \geq 12$) are found within MCs, and these hot materials are heated in the corona by solar flares (Lepri et al. 2001; Lepri & Zurbuchen 2004; Reinard 2005).

SMFRs with durations of tens of minutes were first found and reported by Moldwin et al. (2000), and they proposed that these SMFRs may form as the result of interplanetary magnetic reconnection at the heliospheric current sheet (HCS). Their main evidence for this argument is the absence of reported intermediate-sized MFRs with durations of several hours; if SMFRs and MCs have the same origins, then the size distribution should be continuous. Feng et al. (2007) presented the continuous size distribution of MFRs, which include many intermediate-sized MFRs; they also found that the observed characteristics of MFRs inconsiderably change with the increase in scale. Accordingly, they proposed that, similar to MCs, SMFRs also originate from the Sun and are the interplanetary manifestations of microflare-associated small CMEs, which are difficult to be identified in the coronagraphs. Recently, Feng and Wang (2015) found that SMFRs exhibit the same high-charge-state signatures as MCs, and the hot materials within SMFRs should be heated by related microflare in the corona. Their finding provided reliable evidence for the conjecture that SMFRs and MCs have the same coronal origins, and SMFRs are interplanetary manifestations of small CMEs. In the past ten years, several investigations about SMFRs have been conducted and provided understanding of SMFRs, such as their magnetic and plasma structures, observational properties, origins, and geo-effectiveness (Cartwright & Moldwin 2008, 2010; Feng et al. 2008, 2010a, 2010b, 2011, 2012, 2015; Feng & Wu 2009; Feng & Wang 2015; Gosling et al. 2010; Rouillard et al. 2009, 2011; Tian et al. 2010; Wu, et al. 2009; Zhang et al. 2013; Janvier et al. 2014a, 2014b; Yu et al. 2014, 2016; Zhang & Hu 2016). Many investigators have discussed the origin of SMFRs, and most of them have agreed that SMFRs originate from solar corona. Ordinary CMEs are often associated with prominence eruptions or solar flares, and many of them occur in conjunction with solar flares and prominence eruptions. Accordingly, the low-charge ions can be detected concurrent with the hottest ions within some MCs; the low-charge ions are identified as prominence materials, and the hottest ions are interpreted to be affected by flare heating during the CME initiation (Lepri & Zurbuchen 2010). Compared with ordinary CMEs, the predicted small CMEs may also be frequently associated with small prominence eruptions or/and solar microflare (Feng et al. 2019), and hot materials have been detected within SMFRs. If this deduction is true, then the cool prominence plasma materials can be detected in SMFRs. In this study, we will search for cool prominence materials within SMFRs.

## 2. Method

Solar prominences are arcade-like structures of relatively cool, dense materials suspended in the hotter corona. Hence, the predicted observational characteristics of interplanetary remnant prominence materials should have low temperature, high density, and low ionic charge states, and these signatures can be taken as evidence to identify prominence material. However, the observational low temperature and high density characteristics can be modified by propagation effects when solar wind flows propagate from the Sun to the Earth (Janvier et al. 2014; Feng & Wang 2015; Wang et al. 2018). The new high-density regions can be formed, and original low-temperature regions also can disappear near 1 AU through the non-uniform expansion of magnetic structures in the solar wind (Wang et al. 2018). The ionic charge states in the interplanetary media mainly depend on their solar source temperature because the solar wind ion charge states are nearly frozen near the Sun (Heidrich–Meisner et al. 2016; Wang & Feng 2016; Wang et al. 2018). Thus, the low ionic charge states may be reliable criteria of prominence materials in the interplanetary solar wind, and nearly all the previous reported interplanetary prominence plasma material events exhibit low charge states (e.g., $He^+$, $Fe^{4+}$, $C^{2+}$, $O^{2+}$, $O^{5+}$, $Fe^{5+}$, and $Fe^{6+}$) (Burlaga et al. 1998; Lepri & Zurbuchen 2010; Skoug et al. 1999; Wang et al. 2018). Recently, Wang et al. (2018) took the unusual $O^{5+}$ and $Fe^{6+}$ abundances as indicators to detect prominence materials. They defined unusual $O^{5+}$ abundance as $O^{5+}$ fractions more than or equal to 0.05. The CHIANTI data of Landi et al. (2013) showed that the freezing-in temperature range of the unusual $O^{5+}$ abundance is approximately $1.3$-$3.8 \times 10^5$ K, which is much lower than the classic coronal temperatures. Wang et al. (2018) also defined unusual $Fe^{6+}$ abundance as $Fe^{6+}$ fractions more than or equal to 0.10, and its estimated freezing-in temperature is approximately $0.9$–$3.2 \times 10^5$ $K$. They identified 27 prominence material regions contained within MCs during 1998–2007 using unusual $O^{5+}$ and/or $Fe^{6+}$ abundances. In this study, we will use the unusual $O^{5+}$ and/or $Fe^{6+}$ abundance to detect the possible prominence materials within interplanetary SMFRs.

## 3. Observations

According to the criteria defined by Wang et al. (2018), we examined all the 2-h-averaged $O^{5+}$ and $Fe^{6+}$ fraction data obtained from the Solar Wind Ion Composition Spectrometer (SWICS) on the *Advanced Composition Explorer* (*ACE*) spacecraft during 1998 to 2008, and we found that 13 SMFRs might exhibit low-charge-state signatures of unusual $O^{5+}$ and/or $Fe^{6+}$ abundances. The 13 SMFRs with possible cold material regions are listed in Table 1. The second and third columns show the front and rear boundaries of SMFRs, respectively. The fourth column presents the measured times of the cold materials. The fifth and sixth columns show the $O^{5+}$ and $Fe^{6+}$ fractions of the cold materials. From Table 1 we can find that the values of unusual $O^{5+}$ and/or $Fe^{6+}$ for most events are approaching their thresholds. In addition, The ACE/SWICS data set includes errors and quality flags, and the detection uncertainty is particularly high for these two ions. The enhancement of the $O^{5+}$ and $Fe^{6+}$ fractions on a single data point does not lead to a reliable conclusion that it is associated with cool prominence material. To assess the thresholds of the unusual $O^{5+}$ and $Fe^{6+}$ abundances, we examined the $O^{5+}$ and $Fe^{6+}$ abundances of the solar wind in 2000, and found that unusual $O^{5+}$ and/or $Fe^{6+}$ abundances were observed in 13 time periods. 4 of the 13 time periods were observed within SMFRs (see Table 1), 5 time periods were observed within MCs (see Table 1 of Wang et al. 2018), 4 time periods were observed within ICMEs (ICME on13-15 April, ICME on 1-3 July, ICME on 8-9 September), only one time period was observed before the SMFR on 21 September 2000. This indicates that the unusual $O^{5+}$ and/or $Fe^{6+}$ abundances are usually related to SMFRs or ICMEs (including MCs) MCs, furthermore, all the quality flags for the unusual $O^{5+}$ and/or $Fe^{6+}$ abundances within SMFRs are good. Therefore, the unusual $O^{5+}$ and/or $Fe^{6+}$ abundances within SMFRs might indicate that these SMFRs contain residual prominence materials. In this section, we introduce two SMFRs as examples to exhibit their possible cool prominence materials.

Figure 1 shows the magnetic field data in the geocentric solar ecliptic angular coordinates, the proton speed (V), the proton density ($N_P$), the proton temperature ($T_P$), iron solar wind average charge state $< Q_{Fe} >$, and the $O^{5+}$ and $Fe^{6+}$ fractions during the SMFR on 19 July 2002 passage. The two vertical lines are the front and rear boundaries of the SMFR. The event has the apparent flux rope signatures: the total magnetic field magnitude is enhanced, and the latitude angle $\theta$ of the magnetic field vector within the SMFR decreases slowly from $45^o$ at the front boundary to $-20°$ near the rear boundary. Similar to MCs, SMFRs also can be fitted with constant alpha, force-free, cylindrically symmetric flux ropes (Moldwin et al. 2000; Feng et al. 2007, 2008). Feng et al. (2008) described the fitting method in detail. The observed magnetic fields of the SMFR were fitted with the constant alpha model, and the fitting results reveal that that this helical flux rope is left handed, and its axial direction is ($\theta = 31°$, $\varphi = 234°$). Figure 1 also displays the fitting magnetic field curves (dot lines) based on the constant alpha flux-rope model. The two sets of curves are approximately consistent, which indicates that the model fits the observed data well. For this SMFR, SWICS does not detect the $Fe^{6+}$ ion, but the $O^{5+}$ fraction curve reveals that $O^{5+}$ fraction reaches the maximum value at around 17:77 UT, and the maximum value of $O^{5+}$ fraction is 0.055. According to the selection criteria of Wang et al. (2018), the cold plasma material around 17:47 UT may be remnant prominence materials. Although mean Fe charge state data are not available within the SMFR, the mean Fe charge state curve in Figure 1 shows that high mean Fe charge signatures are observed near the boundaries of the SMFR, and their measured values are more than 12. Since the time resolution of mean Fe charge state data is one-hour, the values of the two data points before the front boundary are very close, and the interval between the nearest measured point and the front boundary is only 5 minutes. So it is quite possible that the mean Fe charge state is more than 12 around the front boundary. The Fe average charge states in normal solar wind are from 9 to 11 (Lepri et al. 2001), a high mean Fe charge state ($<Q_{Fe}> \geq 12$) is the most reliable indicator to identify ICMEs (Lepri & Zurbuchen 2004), and high mean Fe charge states are the result of flare-related heating in the corona (Reinard 2005). These observations reveal that this SMFR unexpectedly contains hot and cold plasma. In other words, the predicted small CME related to the SMFR may be associated with a small prominence eruption and a solar microflare.

Figure 2 displays the SMFR on 21 September 2006. This SMFR is a very interesting event, its duration is only about 2 h, and its maximum magnetic field strength is only 2.9 nT. However, this event demonstrates very good flux rope signatures: the magnetic field direction rotates smoothly, and the total magnetic field magnitude is enhanced. The fitting results show that the magnetic helicity of the SMFR is right handed, and the ACE spacecraft traverses the SMFR close to its magnetic axis. Figure 2 shows that the $Fe^{6+}$ fractions have an apparent enhancement within the SMFR, and the value of $Fe^{6+}$ fraction reaches 0.127 at 18:24 UT. The $O^{5+}$ fraction curves are also enhanced in the corresponding point, and its value is 0.040. As mentioned above, low ionic charge states can also be taken as evidence in the search for prominence material. Thus, the cool plasma near 18:24 UT may be prominence materials. In addition, the mean Fe charge maintains a normal level throughout the SMFR, and its maximum value is 10.5. Therefore, the solar source region of the SMFR may be associated with prominence eruptions only (without flare).

As Wang et al. (2018) pointed out, although their selected threshold criteria are less-restrictive, some cold prominence materials in ICMEs may still have been missed. Figure 3 shows a SMFR on 25 May 2011. The $Fe^{6+}$ abundance curve had an apparent protuberance before the rear boundary, and its maximum value reached 0.096, which is lower but close to the selection criteria of Wang et al. (2018). Therefore, the cold plasma material before the rear boundary may also be remnant prominence materials. In addition, Chi et al. (2018) successfully traced the origin of the SMFR using observations from the Sun–Earth Connection Coronal and Heliospheric Investigation (SECCHI) package on board STEREO. The source region on the solar disk of the progenitor CME of the observed SIMFR has been confirmed at S16W13 on 25 May 2011 by Chi et al. (2018). Thanks to the high quality observations from the SDO/AIA, we have checked the coronal activity in detail. The result indicates that the coronal activity is caused by a small filament (F1; marked by the white arrows in the left panel of Figure 4) eruption, during which F1 underwent a fierce explosion and almost all filament material was ejected from the active region (see the right panel of Figure 4). Since there is no other activity source near the source region within 4 hours before the first observation of the progenitor CME in STEREO/COR2, it can be confirmed that the SMFR is associated with the eruption of F1. And from the observation, we can also verify that the material of F1 was mainly ejected away from the source region during the eruption. Synthesizing these together, we can naturally come to the conclusions that the progenitor CME is a result of the successfully eruption of F1 and definitely includes the material of F1. So the cold plasma material within the SMFR should be remnant prominence materials.

Although these observations indicated that some SMFRs might exhibit low-charge-state signatures of unusual $O^{5+}$ and/or $Fe^{6+}$ abundances. However, ACE/SWICS is not ideally suited to detect $O^{5+}$ and $Fe^{6+}$ reliably, since $O^{5+}$ and $Fe^{6+}$ are close to their detection threshold of ACE/SWICS, and the measured $O^{5+}$ and $Fe^{6+}$ are likely to be contaminated by the tails of other $O$ and $Fe$ ions. In addition, all possible low-charge-state signatures within these SMFRs are only measured at a single data point, the low-charge-state signatures of unusual $O^{5+}$ and/or $Fe^{6+}$ abundances may be caused by instrumental effects. It should be noted that the resolution and accuracy of the available data set does not allow us to rule out that the results are not an instrumental effect.

**4. Summary and Discussion**

SMFRs have two possible origins (Moldwin et al. 2000). One is that SMFRs are small-scale MCs, and they are the interplanetary manifestations of small CMEs. The second possibility is that SMFRs are formed in interplanetary space owing to magnetic reconnections near the HCS. Feng et al. (2007) supposed that SMFRs are manifestations of microflare-associated small CMEs, and Feng and Wang (2015) also found hot materials in SMFRs and proved the conjecture of Feng et al.

(2007). The ordinary CMEs are frequently associated with prominence eruptions, and cool prominence materials are found within some MCs. Thus, we want to know whether small CMEs are frequently associated with small prominence eruptions and whether the cool plasma material can be detected in SMFRs. In this study, we investigated the $O^{5+}$ and $Fe^{6+}$ fractions during 1998–2008 to search for remnant prominence materials within SMFRs, and we found that 13 SMFRs may contain cool plasma materials. The low charge state with MCs indicate an association with prominence eruptions, and our results reveal that some SMFRs might have the same low-charge-state signatures. This would indicate that some SMFRs originated from small CMEs associated with small prominence eruptions. In addition, cool and hot ionic charge states can be detected in some MCs, and their related CMEs are associated with flares and prominence eruptions. Similarly, the SMFR on 19 July 2002 also could exhibit signatures of low and high ionic charge states, namely, its potentially connected small CME might also be associated with flare and prominence eruptions. These observations of possible cool prominence materials within SMFRs not only provide new evidence for the conjecture that SMFRs are small-scale MCs but also imply that the connected small CMEs can be associated with flares and prominence eruptions.

**Acknowledgments:** The authors acknowledge supports from NSFC under grant Nos. 41804162, 41674170. This work is also supported in part by the Plan For Scientific Innovation Talent of Henan Province under grant No. 174100510019. The authors thank NASA/GSFC for the use of data from ACE, these data can obtain freely from the Coordinated Data Analysis Web (http://cdaweb.gsfc.nasa.gov/cdaweb/istp_public/).

**Table 1. The cold materials within small magnetic flux ropes**

| No. | Front boundary [a] | Rear boundary [b] | Start [c] | $O^{5+}$ [d] | $Fe^{6+}$ [e] |
|-----|---|---|---|---|---|
| 001 | 1998/02/14 09:48 | 1998/02/14 12:07 | 1998/02/14 10:49 | 0.031 | 0.122 |
| 002 | 1998/05/05 07:04 | 1998/05/05 12:40 | 1998/05/05 09:07 | 0.010 | 0.102 |
| 003 | 2000/01/18 13:00 | 2000/01/18 15:01 | 2000/01/18 13:44 | 0.020 | 0.145 |
| 004 | 2000/04/15 07:14 | 2000/04/15 09:30 | 2000/04/15 08:47 | 0.015 | 0.107 |
| 005 | 2000/08/09 07:12 | 2000/08/09 10:43 | 2000/08/09 08:39 | 0.057 | 0.058 |
| 006 | 2000/09/21 05:20 | 2000/09/21 06:31 | 2000/09/21 05:37 | 0.052 | 0.000 |
| 007 | 2002/07/19 16:02 | 2002/07/19 22:00 | 2002/07/19 17:57 | 0.055 | 0.000 |
| 008 | 2005/12/09 05:44 | 2005/12/09 12:00 | 2005/12/09 09:57 | 0.037 | 0.116 |
| 009 | 2006/04/07 14:28 | 2006/04/07 19:45 | 2006/04/07 17:20 | 0.050 | 0.000 |
| 010 | 2006/05/02 07:37 | 2006/05/02 13:21 | 2006/05/02 09:43 | Data gap | 0.125 |
| 011 | 2006/09/21 17:45 | 2006/09/21 20:17 | 2006/09/21 18:24 | 0.040 | 0.127 |
| 012 | 2007/01/14 08:09 | 2007/01/14 11:43 | 2007/01/14 08:52 | 0.119 | Data gap |
| 013 | 2008/02/26 11:27 | 2008/02/26 16:17 | 2008/02/26 12:40 | 0.058 | 0.109 |

[a] The front boundary of the small magnetic flux ropes (UT).
[b] The rear boundary of the small magnetic flux ropes (UT).
[c] The measured time of the cold material (UT).
[d] The $O^{5+}$ fraction of the cold material.
[e] The $Fe^{6+}$ fractions of the cold material.

Figure 1. Magnetic field, proton and plasma composition data on 19 July 2002.

Figure 2. Magnetic field, proton and plasma composition data on 21 September 2006.

Figure 3. Magnetic field, proton and plasma composition data on 28 May 2011.

Figure 4. Images from SDO/AIA 304 angstrom pass-band show the source region before (a) and after (b) the filament eruption. The white arrows indicate the active filament and the post eruption arcades in panel (a) and panel (b), respectively.

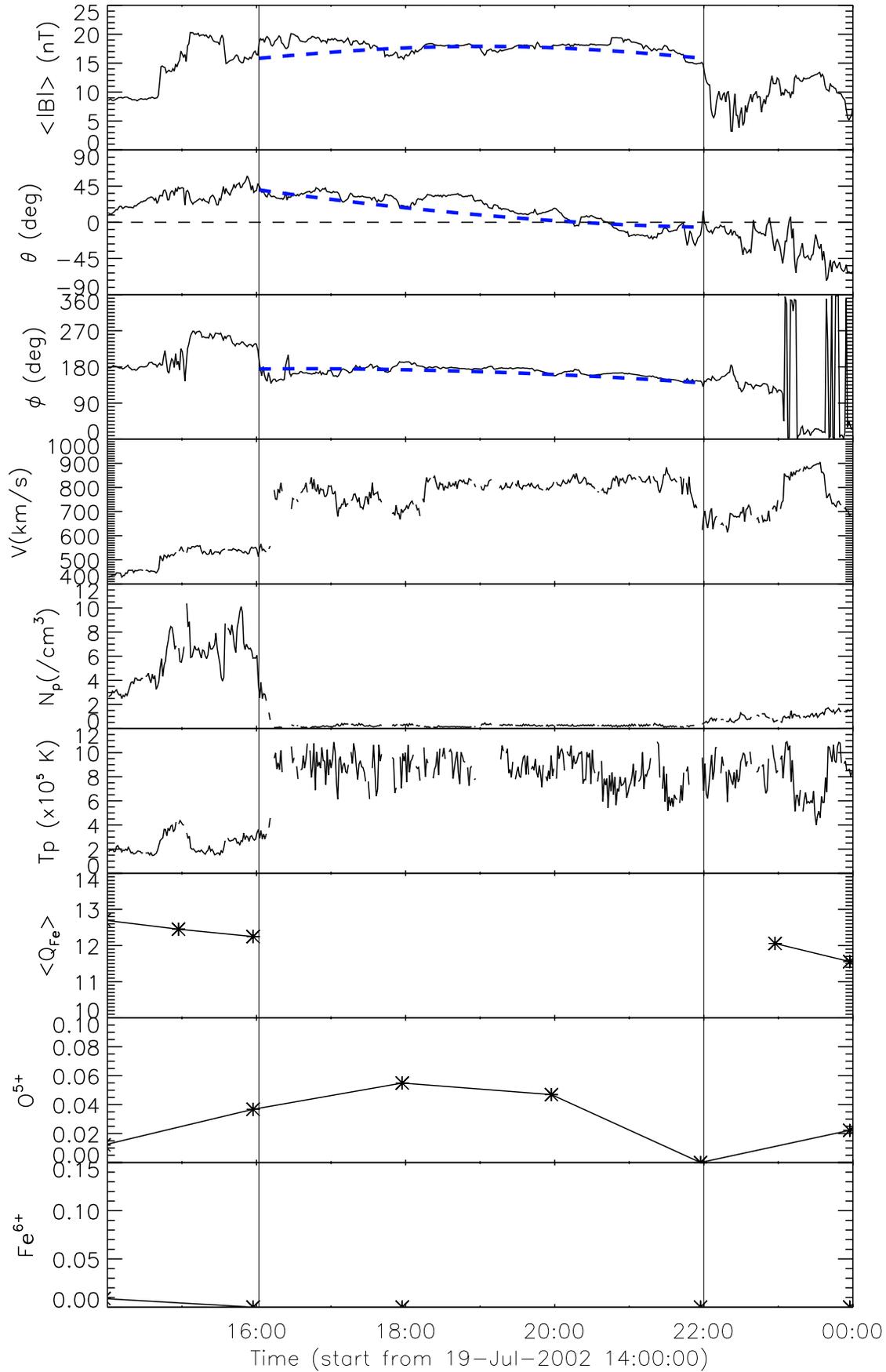

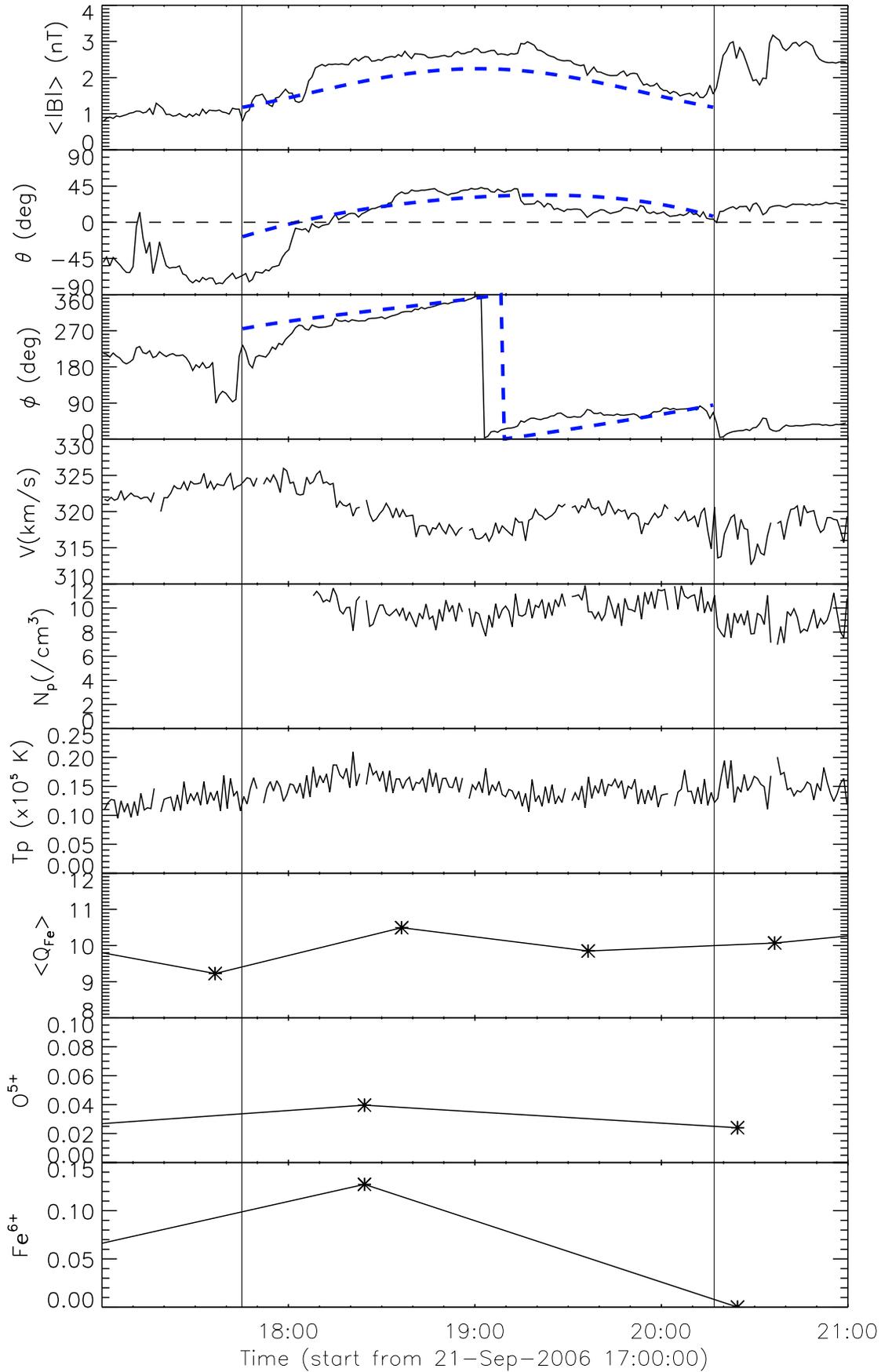

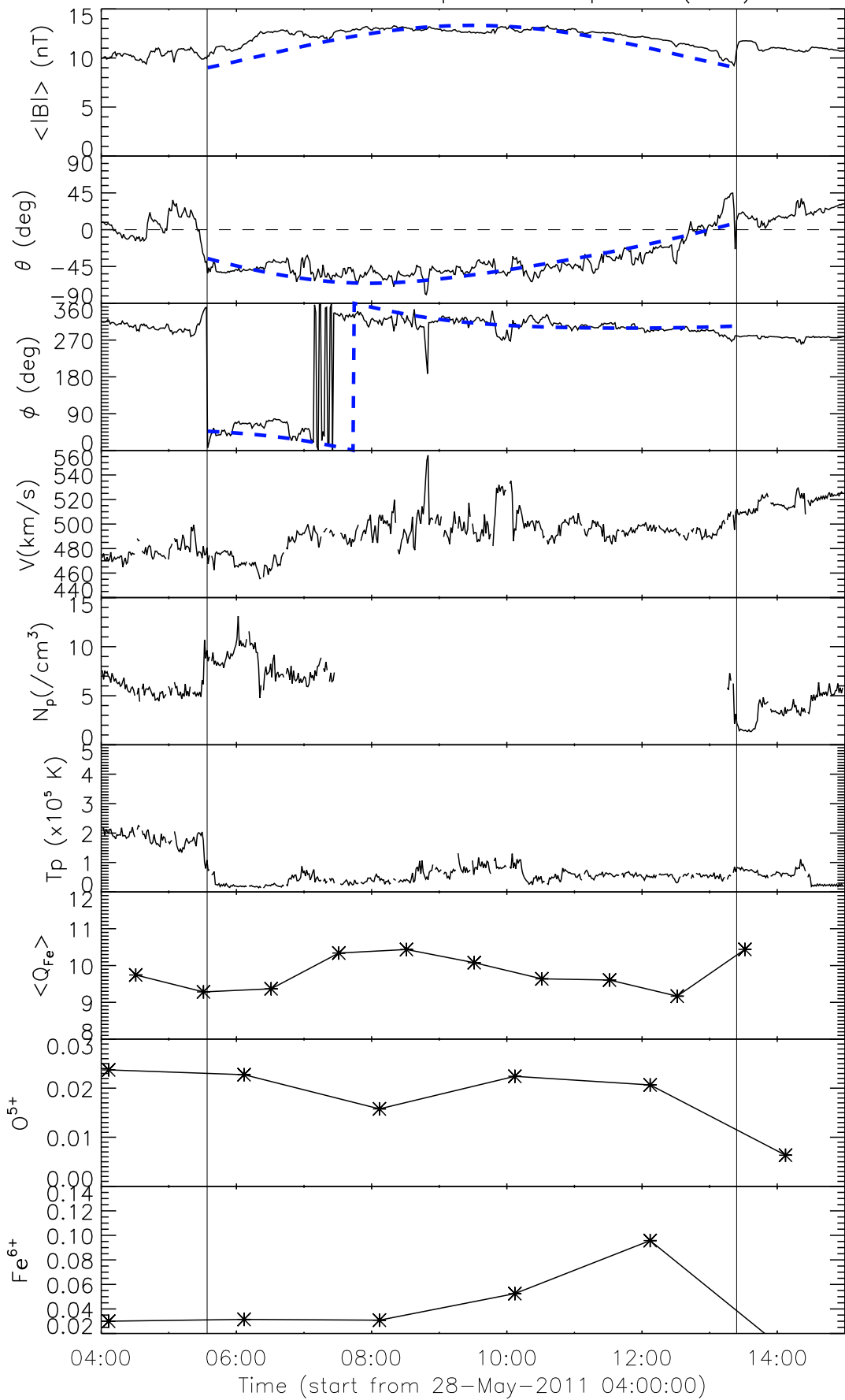

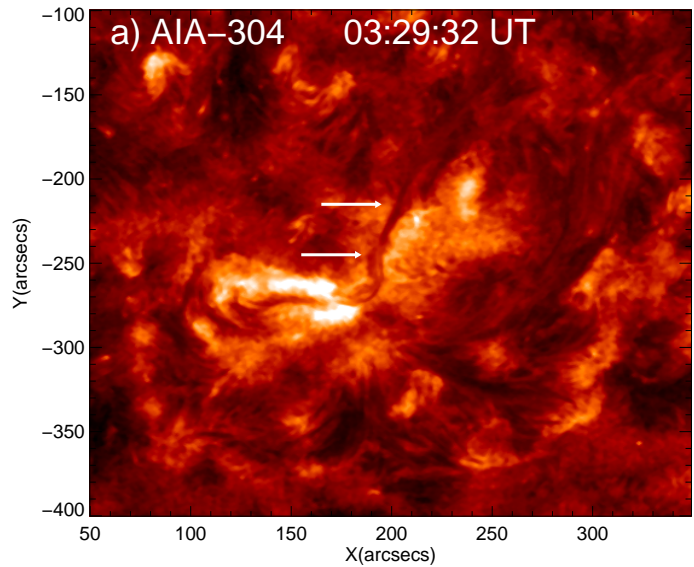 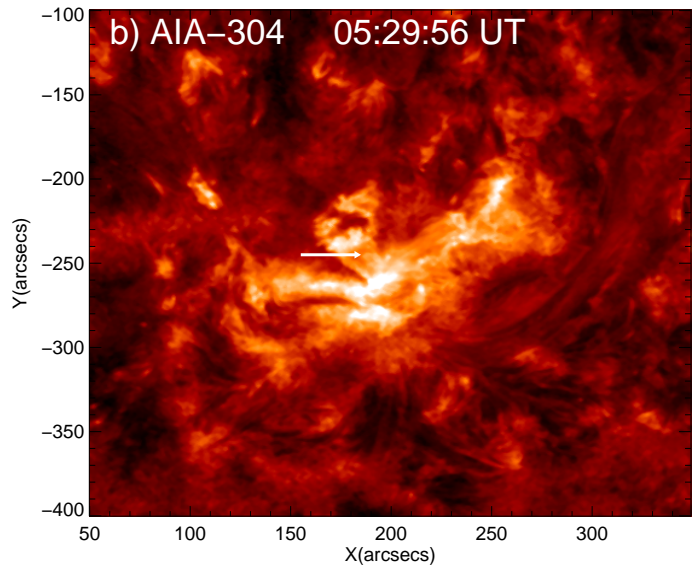